\documentclass{aa}
\usepackage{psfig}
\newcommand{\lsim}{\stackrel{\scriptstyle <}{\phantom{}_{\sim}}}
\newcommand{\gsim}{\stackrel{\scriptstyle >}{\phantom{}_{\sim}}}
\begin{document}

%
   \title{Cooling of Neutron Stars. Hadronic Model.
        \thanks{Research supported in part by the
DFG under grant no. 436 RUS 17/117/03}}


   \author{D. Blaschke \inst{1,2}
        \and H. Grigorian \inst{1,3}
        \and D.N. Voskresensky \inst{4,5}
          }

   \offprints{D. Blaschke}

   \institute{Fachbereich Physik, Universit\"at Rostock,
        Universit\"atsplatz 1, D--18051 Rostock, Germany\\
         email: david.blaschke@physik.uni-rostock.de
         \and Bogoliubov Laboratory for Theoretical Physics,
         Joint Institute for Nuclear Research, 141980 Dubna, Russia
         \and Department of Physics, Yerevan State University, Alex
        Manoogian Str. 1, 375025 Yerevan, Armenia\\
        email: hovik@darss.mpg.uni-rostock.de
        \and Gesellschaft f\"ur Schwerionenforschung mbH, Planckstr. 1,
        D--64291 Darmstadt, Germany
        \and Moscow Institute for Physics and Engineering,
        115409 Moscow, Russia\\
        email: d.voskresensky@gsi.de
             }

   \date{Received: 20. February 2004; accepted month }

\abstract{
We study the cooling of isolated neutron
stars. The main cooling regulators are introduced: EoS, thermal transport,
heat capacity, neutrino and photon emissivity, superfluid nucleon gaps.
Neutrino emissivity includes main processes.
A strong impact of medium effects  on the cooling rates is
demonstrated. With taking into account of medium effects in reaction rates and
in nucleon superfluid gaps modern
experimental data can be well explained.
}
\maketitle

%

\section{Introduction} \label{sec:intro}
The Einstein Observatory was the first that started the experimental study of
surface temperatures of isolated neutron stars (NS).
Upper limits for some sources have been found.
Then ROSAT offered first detections of surface temperatures.
Next $X$-ray data came from Chandra and XMM/Newton.
Appropriate references to the modern data
can be found in recent works by \cite{TTTTT02,T04,KYG01,YGKLP03},
devoted to the analysis of the new data.
More  upper limits and detections are expected from satellites
planned to be sent in the nearest future.
In general, the data can be separated in three groups.
Some data show very {\em{``slow cooling''}} of objects, other demonstrate
an {\em{``intermediate cooling''}} and some show very {\em{``rapid cooling''}}.
Now we are at the position to carefully compare the data with existing
cooling calculations.

\section{Existing NS cooling scenarios}

Let us briefly point out some important achiements
on the way to the present understanding
of the NS cooling problem.
Theoretical study of the NS cooling  has been started long ago in pioneering
works of
\cite{TC65} and \cite{BW65}.
It has been argued that the one-nucleon so called direct
Urca (DU) process, as $n\rightarrow pe\bar{\nu}$, is forbidden
up to sufficiently high density and the main
r\^ole plays the two-nucleon so called modified Urca (MU) process,
like $nn\rightarrow npe\bar{\nu}$ and $np\rightarrow ppe\bar{\nu}$.
As the result of many works the so called
{\em{``Standard scenario''}} of cooling emerged.
It includes the neutrino cooling stage $t \lsim 10^5$~yr and the photon
cooling era, $t \gsim 10^5$~yr.
The MU process and the nucleon-nucleon bremsstrahlung (NB),
as $nn\rightarrow nn\nu\bar{\nu}$ and $np\rightarrow np\nu\bar{\nu}$,
were carefully recalculated  by
\cite{FM79} using the free one-pion exchange.
The emissivity of the most efficient channel of the MU
process $nn\rightarrow npe\bar{\nu}$ is  given by
\begin{eqnarray}\label{MU}
\varepsilon_{\nu}[\rm MU ] \sim
10^{22}~(m^*_{\rm MU}(n))^4\left[\frac{n_e }{n_0}\right]^{1/3}
~\xi_{nn}\xi_{pp}~ T_{9}^8 \,\,\, \frac{{\rm erg}}{{\rm
cm}^{3}~{\rm sec}}~,
\end{eqnarray}
$T_9 =T/10^9 \mbox{K}$, $n_e$ is the electron density and $n_0
\simeq 0.16~fm^{-3}$ is the ordinary nuclear  density at
saturation, $(m^*_{\rm MU}(n))^4  =(m^*_n /m_N)^3 (m^*_p /m_N )$, $m_n^*$ and
$m_p^*$ are effective neutron and proton masses, $m_N$ is the
nucleon mass in vacuum, $n$ is the nucleon density, $(m^*_{MU}(n))^4 (n_e /n_0 )^{1/3}
\sim 0.1$ for $n\sim n_0$.

After \cite{M59} it became clear that NS
in the late cooling stage for $T<T_{cn}$ and $T<T_{cp}$
are neutron and proton superfluids.
The nucleon superfluidity was incorporated in the  {\em{``Standard scenario''}}
by \cite{M79},  who used relevant combinations  of the suppression factors
\begin{eqnarray}
\xi_{ii}\simeq  \mbox{exp}({-\Delta_{ii}/T}), \quad T<T_{ci};
\end{eqnarray}
$\xi_{nn}\xi_{pp}$ for the emissivity of MU $nn\rightarrow npe\bar\nu$
and NB $np\rightarrow np\nu\bar\nu$ processes,
$\xi_{nn}^2$ for NB $nn\rightarrow nn\nu\bar\nu$, $\xi_{pp}^2$ for  MU $np\rightarrow ppe\bar\nu$,
for
$T<T_{cn}$ and $T<T_{cp}$,  $\Delta_{ii}$ is the gap, $i=n$ or $p$.

Typically the NB
emissivity is  by an order of magnitude smaller  than that for MU.
The pre-factor
$\sim 10^{20}\div 10^{21}$ and the temperature dependence
$T_{9}^8$ are typical for the phase space volume of two-nucleon
processes (without inclusion of medium effects, see below).

First cooling calculations were based on the assumption of an isothermal
core. The relevance of the thermal conductivity for the first
$10^2 \div 10^3~$ yr was demonstrated by \cite{NS81}.
A discussion of the achievements of early works can be found in the
review by \cite{T79} and in the book by \cite{ST83}.

The {\em{``Standard scenario''}} allows to explain the {\em{slow cooling}}
but fails to cook the {\em{intermediate cooling}} and
{\em{fast cooling}} of some neutron stars.
To explain the latter different efficient
direct Urca-like processes have been suggested:
the proper DU process, reinvestigated by \cite{LPPH91}
and the DU process going on hyperons (HDU), see the same work;
pion condensation (PU) processes by \cite{MBCDM77};
kaon condensation (KU) processes by \cite{BKPP88,T88}, similar to PU;
and quark DU  processes (QDU) by \cite{I82}. All these processes
show up for baryon densities larger than the corresponding critical ones
$n>n_{c}^{\rm DU}, n_{c}^{\rm PU}, n_{c}^{\rm KU}, n_{c}^{\rm HDU}, 
n_{c}^{\rm QDU} \sim 1.5\div 6 n_0$ according to different model 
dependent estimates. Roughly
\begin{eqnarray}\label{DU}
\varepsilon_{\nu}[\rm DU] &\sim&
10^{27}~(m^*_{\rm DU})^2\left[\frac{n_e}{n_0}\right]^{1/3}T_{9}^6 \nonumber\\
&&\times \mbox{min}[\xi_{nn} ,\xi_{pp} ]\quad
\frac{{\rm erg}}{{\rm cm}^{3}~{\rm sec}}~,
\end{eqnarray}
where $(m^*_{\rm DU})^2 =(m_n^* m_p^* )/m_N^2$.
The factors $10^{26}\div 10^{29}$ and the behavior $\sim T_{9}^6$ are typical
for all mentioned one-nucleon processes.
Thus the {\em{ ``Standard scenario+exotics'' }}
has been considered as a scenario with minimum exotics: some stars have
$n<n_{c}$ and cool slowly whereas some have $n>n_{c}$ in a part of their
interiors and cool very fast.

{\em{However one type of the non-exotic processes was completely
    forgotten already in the ``Standard scenario'' without any
    justification for that.}}
These are the so called neutron pair breaking and formation
(nPBF) and proton pair breaking and formation (pPBF) processes, as they were
named by \cite{SVSWW97}.
The emissivity of the
{\em{nPBF}} process was first calculated by \cite{FRS76} for the
case of the $1S_0$ neutron pairing.
However, their asymptotic expression for the emissivity
$\varepsilon [\mbox{nPBF}]\sim 10^{20}~ T^7_{9}~\exp(-2\Delta_{nn}/T)$
for $T\ll \Delta_{nn}$,
as follows from expression (1b) of their work and from their
rough asymptotic estimate of the
integral (see below (13b)), shows neither the large one nucleon phase space
factor ($\sim 10^{29}$) nor the appropriate temperature behaviour (note that
the full analytic expression of this work is quite correct).
The numerical underestimation of the rate by an order of magnitude
and the very rough asymptotic expression used by \cite{FRS76}
 became the reason that this important result was overlooked for many
years. \cite{VS87} rediscovered the nPBF
process $n\rightarrow n\nu\bar{\nu}$
and introduced the pPBF process $p\rightarrow p\nu\bar{\nu}$
with correct asymptotic behavior of the emissivity
\begin{eqnarray}\label{PFB}
\varepsilon_{\nu} [\mbox{iPBF}]&\sim& 10^{29}~m^*_{\mbox{iPBF}}
\left[\frac{p_{Fi}(n)}{p_{Fn}(n_0)}\right]
~\left[\frac{\Delta_{ii}}{\mbox{MeV}}\right]^{7}~
\left[\frac{T}{\Delta_{ii} }\right]^{1/2}~\nonumber\\
&&\times \xi_{ii}^2 ~~\frac{{\rm erg}}{{\rm cm}^{3}~{\rm sec}}  ~,
\end{eqnarray}
for $T\ll \Delta_{ii}$, $p_{Fi}(n)$ is the neutron/proton Fermi momentum,
$i=n,p$, $m^*_{\mbox{iPBF}}=m_i^* /m_N$.
Eq. (\ref{PFB})
shows a huge one-nucleon phase space factor and
a quite moderate $T$ dependence of the pre-factor
(note that the value of the pre-factor in case of
$1S_0$ neutron pairing evaluated by
\cite{VS87} contains a mistake
that however yields only a non-essential
correction factor of order of one to their final result; a larger uncertainty
comes from not too well known medium modified vertices, see discussion
by \cite{V00} and corrected expressions there).
Medium modifications of vertices (cf. Ward-Takahashi  identity) were 
incorporated in all processes.
This is very important point since medium modified vertices yield an
enhancement factor up to $10^2$ for pPBF compared with the result one would
get using  vacuum vertices.
This factor $\sim 10^2$ arises since the process may occur through $nn^{-1}$
and $ee^{-1}$ correlations, ${-1}$ symbolizes the particle hole,
with subsequent production of $\nu\bar{\nu}$ from the $n$ and $e$ rather than
from a strongly suppressed vacuum vertex of
$p\nu\bar{\nu}$, see \cite{VS87,VKK98,L00,V00}.
The efficiency  of the PFB rates was related by \cite{VS87} to the value of
the pairing gap and it was indicated that these processes can play an
important r\^ole in the cooling scenario.
Note that with taking into account medium effects the estimate (\ref{PFB})
is roughly valid for all the cases for $1S_0$
neutron and proton pairings as well as for the $3P_2$ neutron pairing, except
the specific case, when the projection of the pair momentum onto the
quantization axis is $|m_J |=2$.  In the latter case, where
the gap has zero's in some points of the
Fermi surface, the emissivity would not be exponentially suppressed,
cf. \cite{VS87}.
However conditions, when such a
possibility might be realized, are not clear.

The inclusion of the PBF processes in the cooling
code has first been accomplished by \cite{SVSWW97}
(where also for the first time the
reference to the pioneering work  of \cite{FRS76} appeared)
with the result that variation of gaps
allows to develop a {\em{ ``scenario for intermediate cooling''}}
covering the region of both {\em ``slow and intermediate cooling''}.
Some subsequent works rediscovered the pPBF process
with vacuum vertices thus ignoring medium effects but including
relativistic corrections to the vacuum vertex.
The latter corrections yield an enhancement of order of  $10$ rather than
the mentioned $10^2$ and are therefore of minor importance compared to
the larger contribution of medium effects.
The pPBF emissivity evaluated with vacuum vertex (with relativistic
corrections) was implemented into their cooling code by \cite{YGKLP03}.
Their result for this process should be still enhanced at least
by a factor $\gsim 10$ to account for medium effects.

{\em{Medium effects essentially modify not only the pPBF emissivity
but they correct also the contributions of  all other processes.}}
They have first been included into emissivities of different
processes by \cite{VS84,VS86,VS87}, see also \cite{MSTV90}
and a more recent review \cite{V00}.
It was shown that the main
contribution to the MU process actually
comes from the {\em{pion channel }} of the reaction
$nn\rightarrow npe\bar{\nu}$, where
$e\bar{\nu}$ are radiated from the intermediate pion and the $NN^{-1}$,
exchanging nucleons,
rather than from the nucleon of the leg of the reaction.
Moreover, due to the so called {\em{pion softening}}
(medium modification of the pion propagator)
the matrix elements of the MU process are further
enhanced with the increase of the
density towards the pion condensation critical point.
Thus the corrected
MU process has been called {\em{``the medium modified Urca''}} (MMU)
and corrected NB process was called  MNB.
Roughly, the emissivity
(\ref{MU}) acquires then a factor (mainly due to the pion decay channel
of MMU)
\begin{eqnarray}\label{MMU}
\frac{\varepsilon_{\nu}[\rm MMU ]}{\varepsilon_{\nu}[\rm MU ]} \sim
10^{3}~(n/n_0 )^{10/3}\frac{\Gamma^6 (n) }{{\omega^*}^8  (n)},
\end{eqnarray}
where the pre-factor $(n/n_0 )^{10/3}$ arises from the phase space
volume,
$\Gamma (n)=1/[1+C (n/n_0 )^{1/3}]$, $C\simeq 1.4\div 1.6$,
is the proper nucleon-nucleon correlation
factor, which appears
in the vertices due to the strong interaction,
being
expressed through the Landau-Migdal parameters (in another words $\Gamma (n)$
recovers the Ward-Takahashi identity).
Note that the above
introduced value $\Gamma$
is an averaged quantity. Actually  correlation factors
depend on the
energy-momentum transfer, being different
for vertices connected to the weak coupling ($\Gamma_{w-s}$ is rather
close to unity) and for  vertices related to the pure strong coupling
($\Gamma_{s}$ is slightly less than above introduced factor $\Gamma$,
$\Gamma^6=\Gamma^2_{\rm w-s}\Gamma^4_s$).
${\omega^*} (n)$ is the so called effective pion gap. The latter
has the
meaning of the effective pion mass at finite momentum transfer in
the given reaction, $k=p_{Fn}$. The quantity ${\omega^*}^2  (n)$
replaces the value $(m_{\pi}^2 +p_{Fn}^2 )$ in the case of the free pion
propagator used for the calculation of the MU process by \cite{FM79}.
Thus the squared amplitude of the
nucleon-nucleon interaction contains $\sim \Gamma^2 /{\omega^*}^2 $
instead of $\sim 1/[m_{\pi}^2 +p_{Fn}^2 ]$ for the free one pion exchange used
by \cite{FM79}.
Here we suppressed a smaller contribution of a local
Landau-Migdal interaction, being corrected by $NN^{-1}$ loops,
cf. \cite{VS86,VS87,MSTV90}.
\begin{figure}[htb]
\psfig{figure=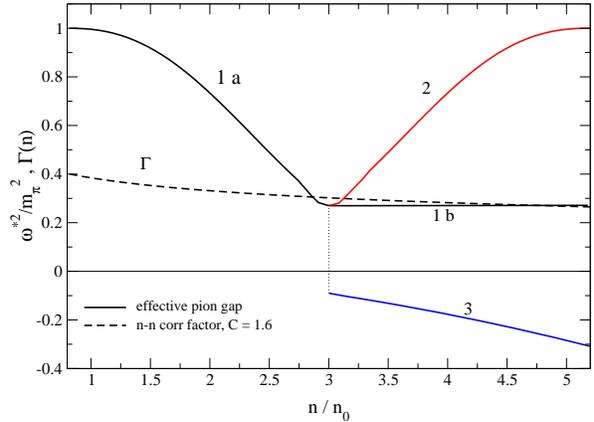,width=0.5\textwidth,angle=-90}
\caption{Nucleon - nucleon correlation factor $\Gamma$ and squared of
the effective pion
gap $\omega^*$ with pion condensation (branches
1a, 2, 3) and without (1a, 1b).
\label{fig1}
}
\end{figure}

The density dependencies of the correlation factor and the effective pion
gap are presented in Fig. \ref{fig1}.
We see that vertices are rather strongly suppressed
(and this suppression increases with the density) but the softening
of the pion mode is enhanced
(${\omega^*}^2 < m_{\pi}^2$) for $n\gsim 0.5\div 0.8 ~n_0$.
Such a behavior  is motivated both theoretically and by analysis 
of nuclear experiments, see \cite{MSTV90,EW88}. 
The curve 1a shows the behavior of the pion gap for $n<n_c^{\rm PU}$.
The value $n_c^{\rm PU}$ depends on different rather uncertain 
parameters (we further assume $n_c^{\rm PU}\simeq 3~n_0$),
cf. discussion by \cite{MSTV90}.
The curve 1b demonstrates the possibility  of a saturation of pion 
softening and the absence of pion condensation for $n>n_c^{\rm PU}$
(this possibility could be realized, e.g., if Landau-Migdal
parameters increased with the density).
Curves 2, 3 demonstrate the possibility of pion condensation for
$n>n_c^{\rm PU}$.
The continuation of the branch 1a  for $n>n_c^{\rm PU}$ (the branch 2)
demonstrates the reconstruction of the pion dispersion relation on the ground
of the condensate state.
Here for simplicity we do not distinguish between
$\pi^0$, $\pi^{\pm}$ condensations, see \cite{VS84,MSTV90}
and the so called alternative-layer-structure,
including both types of condensates, see  \cite{UNTMT94}.
In agreement with a general trend known in condensed
matter physics fluctuations dominate in the vicinity of the critical point
of the phase transition and die out far below and above the critical point
(see the curves 1a, 2).
The jump from the branch 1a to 3 is due to the first order phase transition
to the $\pi$ condensation,
see discussion of this point by \cite{VM82,MSTV90}.
The branch 3 yields the amplitude of the pion condensate mean field for
$n>n_c^{\rm PU}$.   
The observation that the pion condensation appears by the first order
phase transition needs a comment. With the first order phase
transitions in the systems with several charged species is associated 
the possibility of the mixed phase, see \cite{G92}.
The emissivity is increased within the mixed phase since  efficient  DU-like
processes due to nucleon re-scattering on the new-phase droplets are possible.
However \cite{VYT02,MTVTC03} demonstrated that, if exists, the mixed 
phase is probably realized only in a narrow density interval due to
the charge screening effects. 
Thereby to simplify the consideration we further disregard
the possibility of the mixed phase. 
We also disregard the change in the equation of state (EoS) assuming 
that the phase transition is rather weak.

The works of  \cite{VS84}, \cite{VS86} suggested to explain the difference in 
the surface temperatures of various compact objects by the assumption that the
objects  have different masses and thus different density profiles and
different cooling rates, according to above argumentation.

The most precise measurement of NS masses comes from the binary pulsar system 
PSRB1913+16, where for the more massive a value 
$M = 1.4411 \pm 0.0007~M_{\odot}$ has been obtained \cite{TW89}.
The approximate coincidence of this value with the present statistical 
average value for masses of binary pulsars $M\approx 1.35 \pm 0.04 ~M_\odot$
\cite{TC99}, 
motivated some authors to conjecture that all NS masses should be very close 
$1.4~M_{\odot}$. Recent
measurement \cite{L04} of the NS mass $M\simeq 1.25 M_{\odot}$ in JC737-3C3J
does not confirm the above conjecture. NS masses can vary in some limits.
Thus the argumentation, cf. \cite{VS86}, that {\em slow} to {\em  fast}
cooling is explained by different masses of the corresponding objects seems
rather natural.
Many other in-medium reaction channels have been studied.
The particular r\^ole of medium effects in the NS cooling has been reviewed
by \cite{MSTV90} and more recently by \cite{V00}.
The pion softening with increasing baryon density
and the subsequent pion condensation are
now-days reproduced not only by microscopic models of the
polarization of the baryonic medium, see \cite{MSTV90},  and \cite{SST99},
but also by detailed variational calculations of \cite{APR98},
yielding $n_c^{\rm PU}\sim 1.5\div 3~n_0$ for
the $\pi^0$ and charged $\pi^-$ condensates.
We repeat here the conclusion of the series of above mentioned works that
{\em{only due to the enhancement of medium
polarization with the baryon density the pion
condensate may set in and it seems thereby not justified to ignore the
softening effects for  $n<n_{c}^{\rm PU}$, 
and suddenly switch on the condensate for $n >n_{c}^{\rm PU}$.}}

In-medium modifications of different processes including MU, NB,
PFB, PU and DU reactions were implemented in the cooling
code by \cite{SVSWW97} with the conclusion that the data existing
to that time could be explained assuming different masses of the sources.
The statement was demonstrated on different models for the equation of
state (EoS) with and without superfluidity effects.
However, the best fit to the whole set of the data was not
elaborated. E.g., the slow cooling object PSR 1055-52 was not fitted.
As we show below it cannot be done without
a fit of the $1S_0$ n and p
and the $3P_2$ n gaps, their absolute values and density dependencies.
An alternative assumption of an internal heating suggested to explain
a high temperature of this object, although possible, seems us rather 
artificial and we drop it.
Inclusion  of different medium effects (see \cite{V00}) allows  to
develop a general {\em{ ``Nuclear medium cooling scenario''}} describing
all the cooling data and covering the gap between
{\em{ ``Standard''}} and  {\em{ ``Standard+exotics''}} scenarios.
The works of \cite{BKV00} and  \cite{BGV01} considered possibility 
of color superconducting pure quark and  hybrid stars discussing 
peculiarities of these objects.

The newly appeared data were studied in recent works by \cite{KYG01,YGKLP03}
and \cite{TTTTT02,T04}. Medium effects in the emissivity
were disregarded.
The cooling   scenario
based on the inclusion of the DU process ({\em{ ``Standard+exotics DU''}})
has been worked out by \cite{KYG01,YGKLP03}.
Their result is that the neutron $3P_2$ gap should be strongly suppressed
since otherwise one could not explain the ``slow cooling'' objects.
This assumption is supported
by some modern microscopic calculations of the $3P_2$ gap including medium
effects, see \cite{SCLBL96,LS00,SF03}.
Contrary, according to  \cite{KYG01,YGKLP03} the
$1S_0$ proton gap should be enhanced in order to smoothen the
transition from the {\em{``slow cooling''}} to the {\em{ ``rapid cooling''}}.
Only switching on the DU
process for $M>1.358 M_{\odot}$ in their EoS model
allows to
explain the {\em{`` intermediate cooling''}} and {\em{ ``rapid cooling''}}
of some objects.
The transition from {\em{`` slow cooling''}} to
{\em{ ``rapid cooling''}} occurs in a narrow window of masses.

Criticism of the Yakovlev et al.  DU  {\em{``Standard+exotics DU''}}
scenario is as follows, see also \cite{TTTTT02,T04}.
i)
The most elaborated model of the variational theory EoS of \cite{APR98}
permits the DU process only for rather high density $n>n_{c}^{\rm DU}\simeq
5.19~n_0$ and $M\geq 1.839~ M_{\odot}$,  whereas \cite{YGKLP03} used a model
of the EoS, where the DU process starts for $n>n_{c}^{\rm DU}\simeq 2.8 ~n_0$
($M>1.358 ~M_{\odot}$). It seems doubtful that the DU reaction "might
know" that it should start namely, when the NS mass is approximately
$1.4 ~M_{\odot}$, being very close to the average value of the NS mass
measured in NS binaries.  
The NS mass is governed by the full $NN$ interaction whereas the value
of the proton fraction, being responsible for the critical density
of the DU reaction,  is governed
by only a part of the interaction related to the symmetry energy.
In the light of the recent observation of the object with
$M\simeq 1.25 ~M_{\odot}$ the above assumption seems even more unlikely.
If the DU process occurred at $M>1.839 ~M_{\odot}$ as follows from the EoS
by \cite{APR98}, the model of \cite{YGKLP03} based on the opening of the
DU channel would fail.
Indeed, it would mean that the majority of the isolated NS
have masses $M>1.839 ~M_{\odot}$.
The latter sounds quite unrealistic.
ii) \cite{YGKLP03} introduced by hand
a large proton gap. However, there are indications
by  \cite{SCLBL96} that medium effects should suppress the proton gap.
Moreover, \cite{TT97} demonstrated that proton and neutron
superfluidities must be very weak for densities when the DU
process becomes operative.
iii)  As we argued, the pPBF emissivity should be enhanced
by an order of magnitude by medium effects, what is not taken into account.
Moreover, iv) the MU and the NB processes are significantly
affected by medium effects (should be replaced to MMU and MNB) what is
not incorporated.

Works of \cite{TTTTT02,T04} used the {\em{ ``Standard+exotics PU''}}
scenario (based on the inclusion of the PU process
for $n>n_{c}^{\rm PU}\simeq 2.5 ~n_0$ and  $4n_0$ respectively).
In general their {\em{ ``slow cooling''}} curves
are below those of  \cite{YGKLP03}. Thereby
the hottest object PSR 1055-52 (if the
surface temperature of it is correctly extracted from the measurement)
is not described, similar to the result of \cite{SVSWW97}.
The transition from the {\em{ ``slow cooling''}}  to the
{\em{ ``rapid cooling''}} is here due to the PU process.
Again the model assumes that the value of the
transition density for the non-standard process (here pion condensation)
is very close to the value of the central density of a "magic" star
of $M\simeq 1.4~M_{\odot}$ in order to have all NS
masses very near the value  of $1.4~M_{\odot}$. 
The latter seems rather unlikely in the light of the recent
measurement of the NS star mass $M\simeq 1.25~M_{\odot}$,
see \cite{L04}. 
The medium effects are disregarded and thus the above
points iii), iv) remained not corrected.

The necessity to include in-medium effects into the NS cooling problem
is a rather obvious issue.
It is based on the whole experience of condensed matter physics, of the
physics of the atomic nucleus and it is called
for by the heavy ion collision experiments, see \cite{MSTV90,V00,IKHV01,W03}.
Their relevance in the NS cooling scenario was demonstrated by \cite{SVSWW97}.
Below we present cooling calculations of NS based on the
{\em{ ``Nuclear medium cooling scenario''}}, see \cite{V00}.
We remove the above shortcomings and show the relevance
of medium effects on the reaction rates and superfluid gaps
by demonstrating the possibility to fit the whole set of
cooling data available by today.

\section{EoS and structure of NS interior, crust, surface}
\subsection{NS interior}
We will exploit  the EoS of \cite{APR98} (specifically the Argonne $V18+\delta
v+UIX^*$ model), which is based on the most recent models for the
nucleon-nucleon interaction with the inclusion of a parameterized three-body
force and relativistic boost corrections. Actually we adopt a simple analytic
parameterization of this model given by  \cite{HJ99} (later on HHJ).
The latter uses the compressional part with the
compressibility $K\simeq 240$~MeV, and a symmetry energy fitted to the data
around nuclear saturation density, and smoothly incorporates causality at high
densities. The density dependence of the symmetry energy is very important
since it determines the value of the threshold density for the DU process.
The HHJ EoS fits the symmetry energy  to the original  Argonne
$V18+\delta v +UIX^*$ model
yielding $n_c^{\rm DU}\simeq~5.19~n_0$ ($M_c^{\rm DU}\simeq 1.839~M_{\odot}$).

\begin{figure}[htb]
\psfig{figure=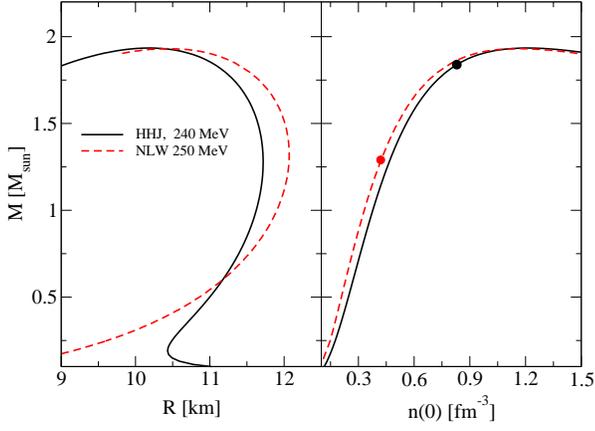,width=0.5\textwidth,angle=-90}
\caption{Gravitational mass-radius relation and mass -central density 
dependence
for the NS configurations corresponding to the HHJ (solid lines) and
NLW (dashed lines) model EoS.
Dots indicate the DU threshold. Possibility of pion condensation
is suppressed. \label{fig2} }
\end{figure}

Fig. \ref{fig2} (left panel) shows the mass-radius relation and (right panel)
the mass-central density relation. Solid lines are for HHJ EoS.
In order to check an alternative possibility and to demonstrate
the sensitivity  of the value of the DU threshold
density  to the selected model
we also use a version of the relativistic
non-linear Walecka (NLW) model in the parameterization of \cite{KV03}.
The parameters of the NLW model are adjusted to the
following bulk parameters of the nuclear matter at saturation:
$n_0 =0.16$~fm$^{-3}$, binding
energy $E_{\rm bind}=-15.8$~MeV,
compression modulus $K=$250~MeV,
symmetry energy $a_{\rm sym}=28$~MeV, and the
effective nucleon mass  $m_N^* (n_0 )=0.8\, m_N$.
The corresponding  coupling constants
are as follows:
\begin{eqnarray}
\label{lagconst2}
\frac{g_{\omega N}^2\,m_N^2}{m_\omega^2}&=&91.2506\nonumber\\
\frac{g_{\sigma N}^2\,m_N^2}{m_\sigma^2}&=&195.597\nonumber\\
\frac{g_{\rho N}^2\,m_N^2}{m_\rho^2}&=&77.4993\nonumber\\
b&=&0.00867497\,,\nonumber\\
c&=&0.00805981\,,
\end{eqnarray}
(note that we treated
obvious misprints in $b$ and $c$ in the paper of \cite{KV03}).
This model was constructed in such a way that it reproduces almost the same
thermodynamic properties as those in the model of \cite{HJ99}.
However, having not enough free parameters all NLW-based models yield
significantly lower threshold densities for
the DU process than those given by variational calculations.
\cite{MSTV90}
treated this fact as a fragile point of the relativistic mean field
models. Thereby they disregarded DU process from their subsequent analysis
concentrating on the pion softening and the PU possibility.
\cite{LPPH91}
used this fact to develop the DU-based scenario for the NS cooling.
Anyhow, in the given NLW model
the threshold density for the DU
process is $n_c^{\rm DU}\simeq~2.62~n_0$ ($M_c^{\rm DU}\simeq 1.29~M_{\odot}$).
Dots in Fig. \ref{fig2} (right panel)
indicate threshold densities for the DU process.
An influence of the pion condensation on the EoS for $n>n_c^{\rm PU}$
is assumed to be small and suppressed thereby.  
For the HHJ EoS the threshold density $n_c^{\rm PU}=3~n_0$ corresponds
to the NS mass $M_c^{\rm PU}\simeq 1.32~M_{\odot}$.
From Fig. \ref{fig2} one can see that deviations in the $M(n)$
relation for HHJ and NLW EoS are minor, whereas the DU thresholds are 
quite distinct.
\begin{figure}[htb]
\psfig{figure=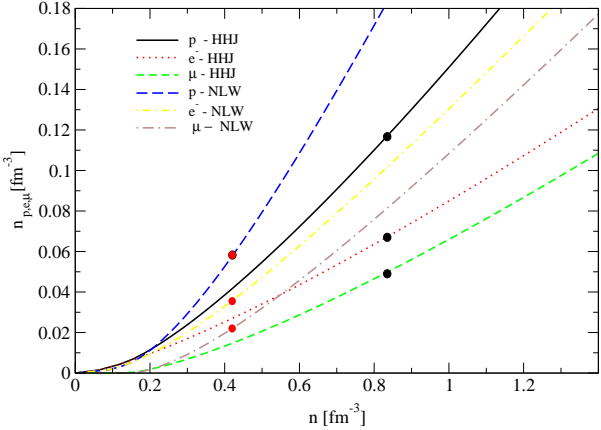,width=0.5\textwidth,angle=-90}
\caption{Densities of the charged particles as a function of
baryon density for HHJ and NLW model EoS. Threshold density for the DU process
is indicated.
An influence of pion condensation for $n>n_c^{\rm PU}$
is neglected.\label{fig3}
}
\end{figure}

Fig. \ref{fig3} demonstrates the concentrations of $p$, $e$ and $\mu^-$ in
HHJ and NLW models as a function of the baryon density.
All these dependencies are quite different for these two models.
This means that  relativistic mean field models may describe well
thermodynamic properties but yield quite different cooling picture compared
with that given by more microscopically based variational calculations of
Argonne-Urbana.
The possibility of charged pion condensation is suppressed.  
Otherwise for $n>n_c^{\rm PU}$ the isotopic composition may change, 
see \cite{MSTV90}, in favor of an increase of a proton fraction and
a smaller critical density for the DU reaction.

\subsection{NS crust}
The density $n\sim 0.5\div 0.7 ~n_0$ is the boundary of the NS
interior and the inner crust.
The latter is constructed of a pasta phase discussed by \cite{RPW83},
see also recent work of \cite{MTVTCM04}.
Then there is the outer crust and the envelope.
Note that our code generates the  temperature profile being inhomogeneous
during first $10^2 \div 10^3~$ yr.
The influence of the crust on the cooling and heat transport is rather minor
basically due to its rather low mass content.  
Thereby the temperature also changes slightly in the crust up to the envelope.

\subsection{Envelope}
Further on we need the relation between the  crust and the surface 
temperature for NS.  
The sharp change of the temperature occurs in the envelope.
This $T_{\rm s} - T_{\rm in}$ relation has been calculated in several works,
see \cite{GS80}, \cite{YLPGC03}, depending on the assumed value of the
magnetic field at the surface and some uncertainties in our knowledge
of the structure of the envelope.
\begin{figure}[htb]
\psfig{figure=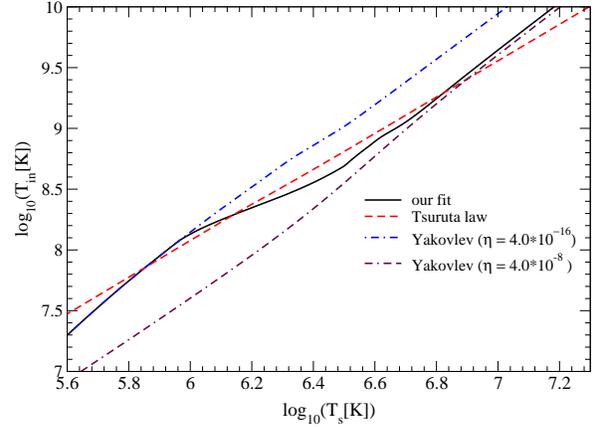,width=0.5\textwidth,angle=-90}
\caption{Relation between the inner temperature and the surface temperature
for different models used in our calculations.
Dash-dotted curves indicate  boundaries of possible values 
$T_{\rm s} =f(T_{\rm in})$, see \cite{YLPGC03}.\label{fig4} }
\end{figure}
Fig. \ref{fig4} shows a range of available $T_{\rm s} =f(T_{\rm in})$ curves,
taken from  \cite{YLPGC03}.
The solid curve, $T_{\rm s} =T_{\rm s}^{\rm fit}=f(T_{\rm in})$,
corresponds to our best fit of the {\em{``slow cooling''}} objects.
It matches the upper boundary (the dash-dotted curve 
$\eta = 4.0 \cdot 10^{-16}$) for rather low $T_s$
(in interval $\log T_{\rm s}[K] \simeq 5.8\div 6$)
and the  lower boundary (the dash-dotted curve $\eta = 4.0 \cdot 10^{-8}$)
for high $T_s$ ($\log T_{\rm s}[K] \geq 6.8$).   
The dash curve shows the so called simplified {\em{``Tsuruta law''}}
$ T_{\rm s}^{\rm Tsur}=(10 ~T_{\rm in})^{2/3}$
used in many old cooling calculations.
We will further vary different possibilities.
The dependence of the results on the $T_{\rm s} - T_{\rm in}$ relation is
demonstrated below in Figs. \ref{fig9} and \ref{fig10}, Figs. \ref{fig12},
\ref{fig13} and \ref{fig14}, and Figs. \ref{fig16}  and \ref{fig17}.

\section{Main cooling regulators}
We compute the NS thermal evolution adopting our fully general relativistic
evolutionary code.
This code was originally constructed
for the description of  hybrid stars by \cite{BGV01}.
The main cooling regulators are the thermal conductivity, the heat capacity
and the emissivity.
In order to better compare our results with results of other groups
we try to be as close as possible to their inputs for the quantities which we
did not calculate ourselves. 
Then we add inevitable changes, improving EoS and including medium effects.

\subsection{Thermal conductivity}
We take the electron-electron contribution to the
thermal conductivity and the electron-proton contribution
for normal protons from \cite{GY95}. 
The total contribution related to electrons is then given by
\begin{eqnarray}\label{kapel}
1/\kappa_e = 1/\kappa_{ee}+1/\kappa_{ep}.
\end{eqnarray}
For $T>T_{cp}$ (normal "n" matter), we have $\kappa_{ep}^{\rm n} =\kappa_{ep}$.
For $T<T_{cp}$ (superfluid "s" matter), \cite{GY95} suggested to drop 
the superfluid contribution $1/\kappa_{ep}^{\rm s}$. 
We use the expression
\begin{eqnarray}
\kappa_{ep}^{\rm s}  = \kappa_{ep}/\xi_p >  \kappa_{ep}^{\rm n} ,
\end{eqnarray}
that gives a crossover from the non-superfluid case to the superfluid case.
The vanishing of $\kappa_{ep}^{\rm s}$ for $T\ll T_{cp}$
is a consequence of the scattering of superfluid protons on the electron
impurities, see \cite{BGV01}.
Following  (\ref{kapel}) we get $\kappa_e^{\rm n}  < \kappa_e^{\rm s}$.
Surprisingly, it is in disagreement with Fig. 4 of \cite{BHY01}, where
probably the curves ``SF1'' and ``SF0'' should be interchanged.

For the nucleon contribution,
\begin{eqnarray}
\kappa_n = 1/\kappa_{nn}+1/\kappa_{np}~,
\end{eqnarray}
we use the result of \cite{BHY01} that includes corrections due to the
superfluidity.
Although some medium effects are incorporated in this work, the
nucleon-hole corrections of correlation terms and the modification
of the tensor force are not included. 
This should modify the result.
However, since we did not calculate $\kappa_n$ ourselves, we
may only roughly estimate the modification.
As we have shown above in Fig. 1, not too close to the critical point of the
pion condensation the squared matrix element of the
$NN$ interaction  $|M|^2_{\rm med} \sim p_{F,n}^2
\Gamma^2 /\widetilde{\omega}^2$
is of the order of the corresponding quantity
$|M|^2_{\rm vac} \sim p_{F,n}^2 /[m_{\pi}^2 + p_{F,n}^2 ]$
estimated with the free one pion exchange, whereas $|M|^2_{\rm med}$
may significantly increase for $n\sim n_c^{\rm PU}$.
To simulate the effect we just allow for the variation of $\kappa_n$
multiplying it by the factor  $\zeta_{\kappa}=10$ (fast transport)
and  $\zeta_{\kappa}=0.3$ (slow transport), see 
Figs. \ref{fig7}, \ref{fig18} below.
The former case is relevant for  rather massive stars, whereas the latter,
for rather light stars.
A suppression of the nucleon-nucleon amplitude compared to the one for the
free one pion exchange model for
$n\lsim n_0$ is motivated by the in-medium $T$ matrix calculations by
\cite{BRSSV95}.

The total thermal conductivity is the straight sum of the partial
contributions
\begin{eqnarray}
\kappa_{tot} = \kappa_{e}+\kappa_{n}+...
\end{eqnarray}
Other contributions to this sum are
smaller than those presented explicitly ($\kappa_e$ and $\kappa_n$).

\subsection{Heat capacity}
The heat capacity contains nucleon, electron, photon, phonon, and
other contributions.
The main in-medium modification of the nucleon heat capacity is due to the
density dependence of the effective nucleon mass. We use the same expressions
as \cite{SVSWW97}.
The main regulators are the nucleon and the electron contributions.
For the nucleons ($i=n,p$), the specific heat is (\cite{M79})
\begin{equation}
c_i \sim 10^{20}({m_i^*}/{m_i})~(n_i/n_0)^{1/3} \zeta_{ii}~T_9~
{\rm erg~cm^{-3}K^{-1}}~,
\end{equation}
for the electrons it is
\begin{equation}\label{e}
c_e \sim 6\times 10^{19} \,(n_e/n_0)^{2/3}~T_9~
{\rm erg~cm^{-3}~K^{-1}}~.
\end{equation}
Near the phase transition
point the heat capacity acquires a fluctuation contribution. For the
first order pion
condensation phase transition this additional contribution contains no
singularity, in difference
with what would be for the second order phase transition, see
\cite{VM82,MSTV90}. Finally, the nucleon contribution to the heat capacity
may increase up to several times in the vicinity of the pion condensation
point. The effect of this correction on global cooling properties is rather
unimportant.

The symmetry of the
$3P_2$ superfluid phase allows for the contribution of
Goldstone bosons (phonons):
\begin{eqnarray}
C_{G}\simeq 6\cdot 10^{14}T_9^3 \,\,\,
\frac{{\rm erg}}{{\rm cm}^{3}~{\rm K}},
\end{eqnarray}
for $T<T_{cn}(3P_2)$, $n>n_{cn}(3P_2)$. We also
include this contribution in our
study although its effect on the cooling is rather minor.

\subsection{Emissivity}
We adopt the same set of partial emissivities as in the
work of  \cite{SVSWW97}. The phonon contribution to the emissivity
of the $3P_2$ superfluid phase is negligible.
The main emissivity regulators are the MMU, see above rough
estimation (\ref{MMU}),
nPBF and pPBF processes, see above rough estimation (\ref{PFB}).

Only qualitative behavior of the interaction shown in  Fig. \ref{fig1}
is motivated by microscopic analysis whereas actual
numerical values of the correlation parameter and the pion gap
are rather uncertain.
Thereby we vary the values $\Gamma (n)$ and ${\omega^*}^2 (n)$
in accordance with our above discussion of Fig. \ref{fig1}.
By that we check the relevance of alternative possibilities:
a) no pion condensation and a saturation of the pion
softening with increasing density, b) presence of pion condensation.

We also add the contribution of the DU for $n>n_c^{\rm DU}$, see above
rough estimation
(\ref{DU}).

All emissivities are corrected by correlation effects.
The PU process contains an extra $\Gamma_{s}^2$ factor compared to the DU
process.
Another suppression of PU emissivity comes from the fact that it is
proportional to the squared pion condensate mean field $|\varphi|^2$.
Near the critical point $|\varphi|^2 \sim 0.1$
increasing with the density up to $|\varphi|^2 \sim f_{\pi}^2 /2$, where
$f_{\pi}\simeq 93~$MeV is the pion decay constant.
Finally, the PU emissivity is about 1-2 orders of magnitude suppressed
compared to the DU one.
Moreover, we adopt the same gap  dependence for the PU process as that for the
DU process.

\begin{figure}[htb]
\psfig{figure=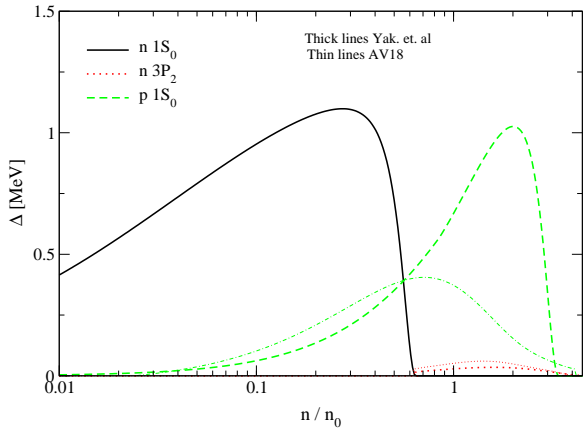,width=0.5\textwidth,angle=-90}
\caption{
Neutron and proton pairing gaps according to \cite{YGKLP03} (thick solid,
dashed and dotted lines) and to \cite{TT04}
(thin lines).
\label{fig5} }
\end{figure}

\subsection{Nucleon superfluidity}

In spite of many calculations,
values of nucleon gaps in dense NS matter are poorly known. This is
the consequence of the exponential dependence of gaps on the
poorly known $NN$ interaction for $n\neq n_0$.
Recent calculations
of \cite{SF03}  who included medium effects in evaluation of
the $3P_2$ gap demonstrate  its strong suppression up to values $\lsim 10~$keV.
Together with previous findings of \cite{SCLBL96,LS00}
they motivate to consider
possibility of rather suppressed gaps.
The suppression of the $3P_2$ gap
for $n>n_c^{\rm PU}$ (for charged $\pi$ condensation),
as explained
in text of the work of \cite{SVSWW97}, is also in coincidence with the
argumentation of \cite{TT97}, see also \cite{TTTTT02,T04}.

Below we start with the model used by \cite{YGKLP03}.
Also we use the gaps from recent work \cite{TT04} to compare the results.
The gaps are presented in Fig. \ref{fig5}.  
Thick lines, the gaps from   \cite{YGKLP03}
and thin lines, from  \cite{TT04}, the model AV18 by \cite{WSS95}. 
The $1S_0$ neutron gap is taken the same in both models.
We allow for a variation of gaps in wide limits
in order to check the sensitivity of the results to their values.
As we will see the cooling curves are sensitive to the magnitudes and
to the density dependence of the gaps.
Therefore further microscopic studies of the gaps are required.

\subsection{KDU, HDU, QDU and other}
The phase structure of dense NS matter might be very rich, including
$\pi^0$, $\pi^{\pm}$ condensates and $\bar{K}^0$, $K^{-}$ condensates
in both $S$ and $P$ waves  (\cite{KV03}); charged $\rho$-meson condensation
by (\cite{V97}); coupling of condensates  \cite{UNTMT94};
fermion condensation yielding an efficient DU-like process
in the vicinity of the pion condensation point (with the emissivity
$\varepsilon_{\nu}\sim 10^{27}~T_9^5$, $m^*_N \propto 1/T$), see \cite{VKZC00};
hyperonization, see \cite{TT04}; quark matter with different phases,
like so called 2SC, CFL, CSL, plus their interaction with meson condensates,
see \cite{RW00,BGV01} and refs therein; and different mixed phases.
In the present work, we suppress all these possibilities
of extra efficient cooling channels.
They are effectively simulated by our PU choice.
The quark matter effects need a special discussion. 
We will return to the latter possibility  in a subsequent publication.

\section{Numerical results}

\subsection{Cooling of normal NS}
Although we have no doubts about the presence of nucleon superfluidity
in NS interiors, we consider first the case of the complete absence
of nucleon superfluidity. The reasons for that are as follows:
i) Thus we  compare our results with previous calculations.
ii) On this example we select more essential and less essential ingredients
in this many-parameter problem. Results are more
transparent being demonstrated on  a simplified example
since the general case introduces new uncertain parameters and
is more involved.
iii) The actual values of gaps might be essentially
smaller than those estimated in the literature, cf. \cite{LS00,SF03}.
This is because most
calculations did not include a proper medium dependence of the $NN$
interactions. Thus we discuss the limiting case of largely suppressed gaps.
\begin{figure}[htb]
\psfig{figure=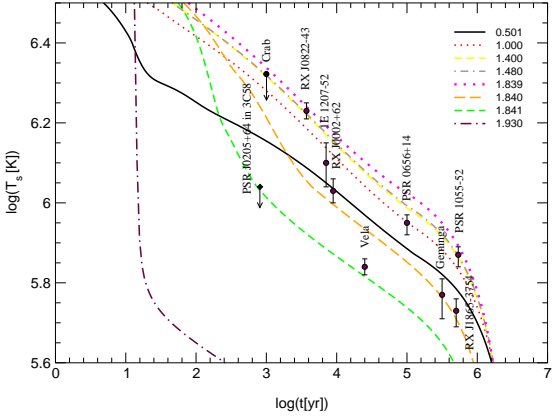,width=0.5\textwidth,angle=-90}
\caption{Cooling of HHJ configurations without medium effects and PU
for
different masses of NS, $T_{\rm s} - T_{\rm in}$ relation
according to our fit. \label{fig6}
}
\end{figure}

Fig. \ref{fig6} demonstrates the cooling evolution 
(for $T_{\rm s}^{\rm fit} (t)$
dependence) of a normal NS for the HHJ EoS. Medium effects  and PU possibility
are disregarded ({\em{``Standard scenario"}}).
We see that one could easily explain the {\em ``slow cooling''} points.
Curves for NS masses in the range
$M\simeq (1\div 1.839)~ M_{\odot}$ lie in this region. 
For $M\simeq 1.839 M_{\odot}$ the efficient DU process is switched on 
within the HHJ EoS ({\em{Standard + exotics DU"}} scenario).
Thereby, the curve corresponding to $M\simeq 1.841M_{\odot}$ jumps down
and already explains the {\em ``rapid cooling''} points.
The  {\em ``intermediate cooling''} points can be explained either
by a very low NS mass (see curve for $M\simeq 0.5~M_{\odot}$, or by
$M\simeq 1.840~M_{\odot}$.
However, it seems rather unrealistic  that
4 from 10 objects either relate to very low
masses like $M\simeq 0.5~M_{\odot}$, or are highly massive, as
$M\simeq
1.840 ~M_{\odot}$, where the DU process is switched on.
{\em ``Rapid cooling''} points are explained by very massive objects
($M\simeq 1.841 ~M_{\odot}$) with DU process. 
They could be also explained by very low-mass objects ($M\sim 0.1~M_\odot$).   
We further drop the latter possibility as rather unrealistic one and 
consider $M\geq 0.5~M_\odot$.

Thus the picture as a whole looks unsatisfactory.
{\em{Only  {\em ``slow cooling''} data are appropriately explained in
a reasonable NS mass interval}} $1.0\div 1.839~M_\odot$. The explanation of 
{\em "intermediate cooling"} and {\em "rapid cooling"}, although
possible, needs very unnatural assumptions.
\begin{figure}[htb]
\psfig{figure=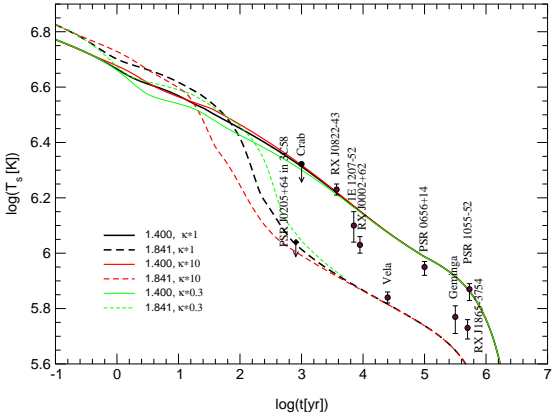,width=0.5\textwidth,angle=-90}
\caption{The influence of the heat conductivity
on the scenario of  Fig. \ref{fig6}.
Two representative configurations with
masses $M=1.40~M_\odot$ (solid lines) and $M=1.841~M_\odot$ (dashed lines).
We vary the heat conductivity by factors
$\zeta=10$ and $\zeta=0.3$.\label{fig7} }
\end{figure}

\begin{figure}[htb]
\psfig{figure=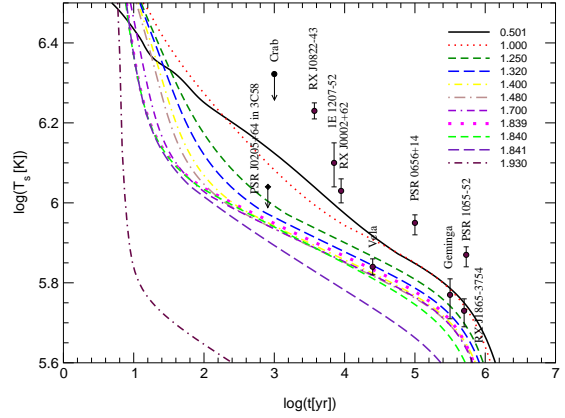,width=0.5\textwidth,angle=-90}
\caption{Cooling evolution of the NS with normal matter HHJ EoS  
(for $T^{\rm fit}_{\rm s}$)
including the medium modification of MU and other processes (MMU, MNB, etc.),
without pion condensation.\label{fig8}
}
\end{figure}
\begin{figure}[htb]
\psfig{figure=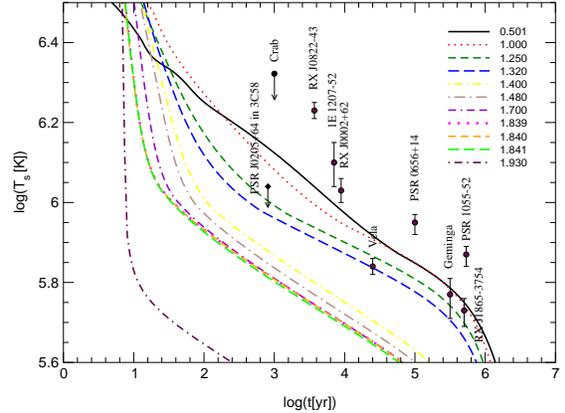,width=0.5\textwidth,angle=-90} \caption{
Same as Fig. \ref{fig8}, including pion condensation for $n>3~n_0$.\label{fig9}
}
\end{figure}
\begin{figure}[htb]
\psfig{figure=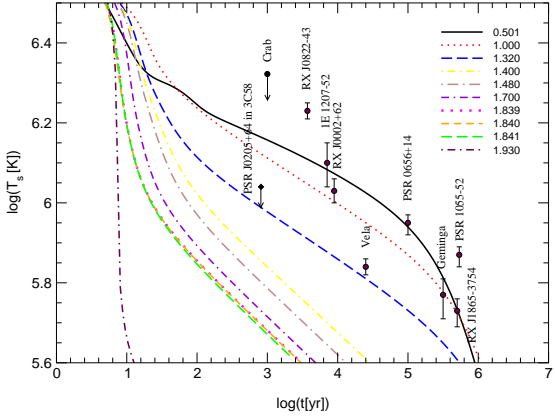,width=0.5\textwidth,angle=-90}
\caption{\label{fig10}
Same as Fig. \ref{fig9}, but with $T_{\rm s} - T_{\rm in}$ relation given by
 the Tsuruta law.}
\end{figure}
Fig \ref{fig7} demonstrates the sensitivity of the cooling
curves to the variation
of $\kappa_n$.
We scale $\kappa_n$ by a factor $\zeta =10$  and by $\zeta =0.3$.
The former case is meaningful for heavy objects whereas the latter one for
low-mass objects.
We see that {\em{both increasing and decreasing of the thermal
conductivity does not change the picture as the whole,}} as well as
the conclusion drawn above.
Transport is relevant only up to the first $ 10^3$~y when the details
of
the $T_{\rm s}-T_{\rm in}$ relation as well as temperature and density
dependences of the cooling regulators determine the response of the 
cooling curves to a rescaling of the heat conductivity by the factor 
$\zeta$.

Fig. \ref{fig8} allows for medium effects the strength of which is
assumed to be saturated with increasing density
(no $\pi$ condensation for $n>3~n_0$). Thus
we use curves 1a, 1b of Fig 1.
We see that medium effects being included in calculation of the emissivity
(MMU, MNB and others) significantly decrease all the curves.
This allows us to explain reasonably well
{\em ``rapid cooling''} points but
{\em ``slow cooling''} and  {\em ``intermediate cooling''}
cannot be addressed without inclusion of superfluidity.   A partial
suppression of
medium effects by scaling of $\omega^*$ does not allow to
improve the picture.

Comparison of Fig. \ref{fig8} and Fig. \ref{fig6} shows that inclusion
of medium effects regulates the mass dependence of the curves. 
In  Fig. \ref{fig6} the curves rise with increase of the NS
mass (for objects with $M\leq 1.89~M_{\odot}$, below DU threshold)
whereas in Fig. \ref{fig8} the trend is changed to opposite.

Fig. \ref{fig9} additionally allows for $\pi$ condensation for $n>3n_0$
($M> 1.32~M_{\odot}$ for HHJ EoS).
Now we use curves 1a, 2, 3 of Fig 1.
The picture remains almost the same as that shown in Fig. \ref{fig8}.
However objects with $M\geq 1.32~M_{\odot}$ are cooled still faster 
by the efficient PU reaction.  
If we assumed $n_c^{\rm PU} \simeq 2.5~n_0$ the pion condensation would start
for NS masses  $M> 1.08~M_{\odot}$.

Fig. \ref{fig10} demonstrates the same cooling evolution of normal NS as
Fig. \ref{fig9}, but for the Tsuruta relation $T_{\rm s}^{\rm Tsur} (t)$.
One can explain the {\em ``rapid cooling''} and the
{\em ``intermediate cooling''} but cannot explain the {\em ``slow cooling''}.
We checked that the use of other  $T_{\rm s} - T_{\rm in}$ relations
in the frame given by Fig. \ref{fig4} does not change general trends.
{\em{Medium effects and the possibility of $\pi$ condensation, although they
regulate the NS mass-dependence show too rapid cooling.}}
We see that without superfluidity the picture is unsatisfactory.
Thereby we conclude that {\em{the cooling data call for the nucleon
superfluidity.}}

\subsection{Cooling of superfluid NS}
First we check the best fit of the model \cite{KYG01,YGKLP03},
where gaps are given by  Fig. \ref{fig5} (thick lines).  
In order to get a fit of the data  within their {\em "Standard + DU"} scenario
\cite{KYG01,YGKLP03} were forced to additionally switch off the $3P_2$
neutron gap.
Since the full switching off the $3P_2$ gap is not supported by
microscopic calculations we simulate the same effect by
introducing
the scaling factor $0.1$ for the $3P_2$ neutron gap shown in  Fig. \ref{fig5}
by corresponding solid line. Then the magnitude of the $3P_2$ gap
becomes to be comparable with the value following from the findings of
\cite{SF03}, who estimated the $3P_2$ neutron gap including medium effects.

\begin{figure}[htb]
\psfig{figure=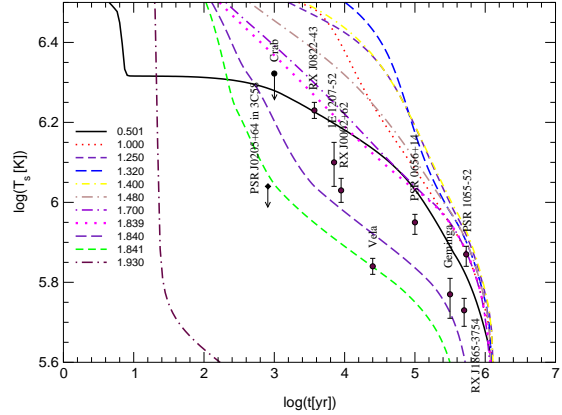,width=0.5\textwidth,angle=-90}
\caption{
Cooling of NS configurations with superfluid nuclear matter without medium
effects and pion condensation. The gaps are taken as in
\cite{YGKLP03}, see Fig. \ref{fig5}, the neutron $3P_2$ gap is additionally
suppressed by a factor $10$, $T_{\rm s} - T_{\rm in}$ relation is
given by $T_{\rm s}^{\rm fit} (t)$.
\label{fig11}}
\end{figure}
Fig. \ref{fig11} demonstrates the cooling evolution ($T_{\rm s}^{\rm fit} (t)$
dependence) of the above choice of gaps for a NS with HHJ EoS.
Medium effects  are not included.
The curves rise compared to those in Fig. 6.
We see that for $M\leq  1.839 ~M_{\odot}$ when the DU process is switched off,
all the curves demonstrate very slow cooling even not explaining the
{\em ``slow cooling''} points (for $M\gsim 1~M_{\odot}$).
We also checked the artificially suppressed rate of the pPBF process
used by \cite{KYG01,YGKLP03}.
Suppression of the rate by a factor of $10$ does not significantly
affect the curves and does not change the conclusion.
The general trends of the curves are similar to those of \cite{KYG01}.
However within the HHJ EoS the {\em ``intermediate cooling''} points and
{\em ``rapid cooling''} points can be explained only by objects with
$M> 1.839 ~M_{\odot}$ and DU cooling what seems us quite unsatisfactory.
Comparison of Fig. \ref{fig11} with  Fig. \ref{fig6}, computed within
the same scenario but without inclusion of the superfluidity, shows 
that both choices suffer of the very same shortcomings.
{\em The data call for medium effects.}

\begin{figure}[htb]
\psfig{figure=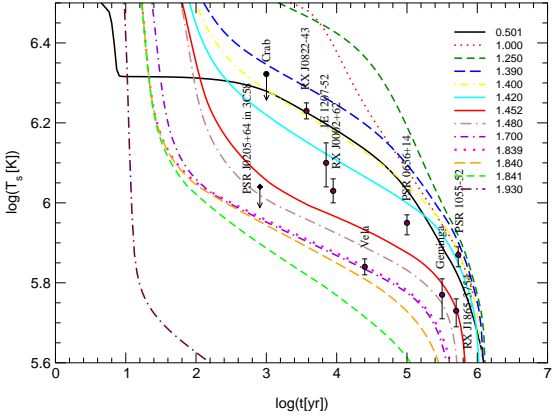,width=0.5\textwidth,angle=-90}
\caption{
Same as in Fig. \ref{fig11} including medium effects without pion
condensation (curves 1a, 1b in Fig. \ref{fig1}).
\label{fig12}
}
\end{figure}

\begin{figure}[htb]
\psfig{figure=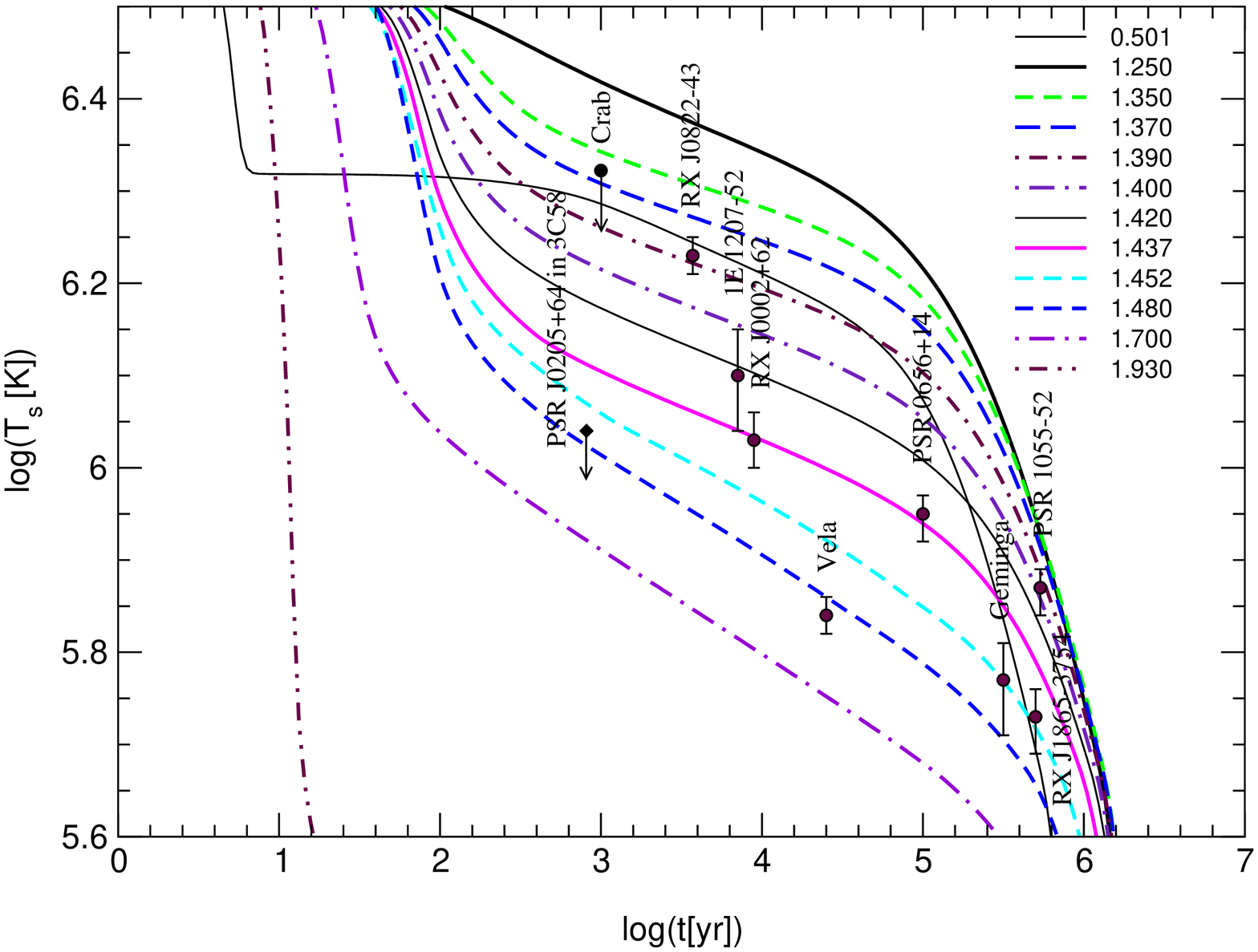,width=0.5\textwidth,angle=-90}
\caption{
Same as in Fig. \ref{fig12} but using the Tsuruta law.
\label{fig13}
}
\end{figure}

\begin{figure}[htb]
\psfig{figure=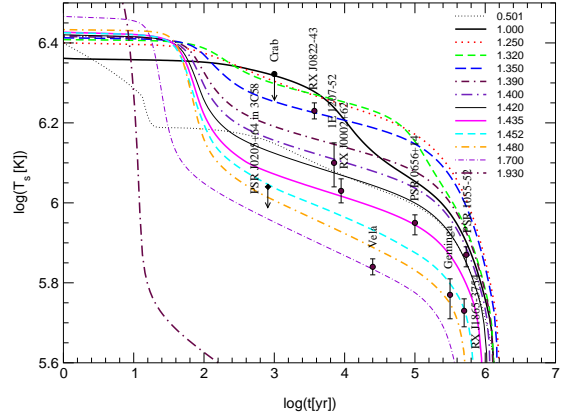,width=0.5\textwidth,angle=-90}
\caption{
Same as in Fig. \ref{fig12} but $T_{\rm s} - T_{\rm in}$ relation
according to \cite{YGKLP03},
$\eta=4\times 10^{-16}$.
\label{fig14}
}
\end{figure}

Fig. \ref{fig12} allows for the medium effects (we use curves 1a, 
1b of Fig \ref{fig1}, no  $\pi$ condensation for $n>3~n_0$).
We smoothly cover all data points.   
All the data are explained by masses in the interval 
$M\simeq 1.39\div 1.84 M_{\odot}$. 
The regular behavior  is clearly seen.
More massive objects cool faster than less massive ones. 
The star with the mass $1.25~M_{\odot}$ may relate to the old and 
hot objects like PSR 1055-52.

We checked a sensitivity of the result to the dependence
$T_{\rm s} - T_{\rm in}$. Fig. \ref{fig13} uses the ``Tsuruta law'' whereas
Fig. \ref{fig14} probes  the choice of  \cite{YGKLP03}, 
$\eta=4.0\times 10^{-16}$.
We see that a variation of  $T_{\rm s} - T_{\rm in}$ does not change the
picture as the whole. 
Only the interval of masses that cover the data is slightly changed.
This interval is  $M\simeq 1.36\div 1.50 M_{\odot}$ according to  
Fig. \ref{fig13} and  $M\simeq 1\div 1.75~ M_{\odot}$ according to  
Fig. \ref{fig14}.
\begin{figure}[htb]
\psfig{figure=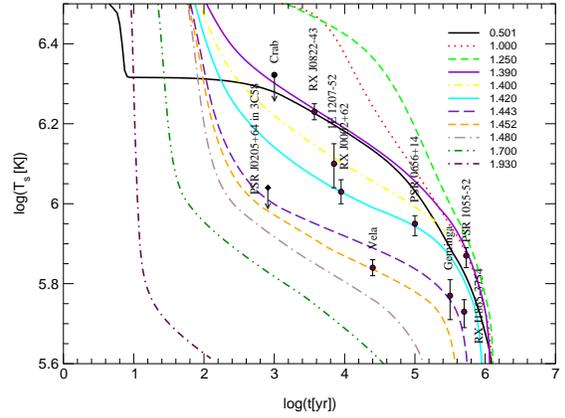,width=0.5\textwidth,angle=-90}
\caption{
Same as in Fig. \ref{fig12} including $\pi$ condensation.
\label{fig16}
}
\end{figure}

\begin{figure}[htb]
\psfig{figure=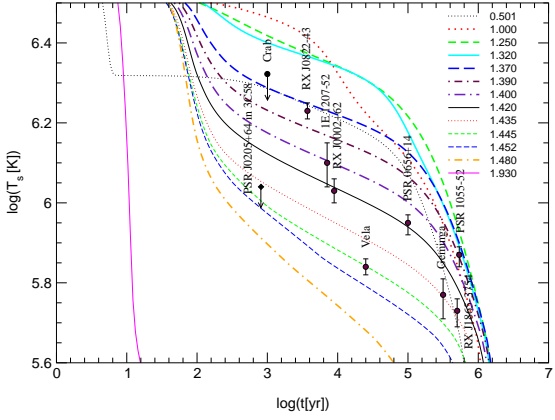,width=0.5\textwidth,angle=-90}
\caption{
Same as in Fig. \ref{fig16} using the Tsuruta law.
\label{fig17}
}
\end{figure}

The mass interval that covers the data is $M\simeq 1.37 \div 1.46~M_\odot$.
Fig. \ref{fig16} allows for $\pi$ condensation for $n>3~n_0$.
Now we use curves 1a, 2, 3 of Fig \ref{fig1}.  
The   $T_{\rm s} - T_{\rm in}$ relation is according to our fit.

Fig. \ref{fig17} shows the same as Fig. \ref{fig16} but for the Tsuruta law. 
In both cases the picture remains almost the same as that shown in  
Fig. \ref{fig13} and Fig. \ref{fig12}, respectively.  
The difference starts for masses $\geq 1.32~M_{\odot}$ due to
switching on the efficient  PU process.
The interval of masses that cover the data is  
$M\simeq 1.36 \div 1.452 ~M_\odot$.
Using of other  $T_{\rm s} - T_{\rm in}$ dependences does not change the 
picture.
Thus we may conclude that the $\pi$ condensation does not contradict
the data but, on the other hand, this assumption is not motivated by 
the existing cooling data.
The hyperon enhanced cooling or kaon condensation cooling do not change the
picture if their critical densities are $\gsim 3~n_0$.  
Note  that the transition density can't be too low.  
At least $n_c^{\rm PU} \geq 2.5\div 2.7~n_0$ for the charged
pion condensation within given EoS.  
The neutral PU process might be additionally suppressed compared to
the charged one, see \cite{V00}, that may weaken the above restriction.
Otherwise we would get too fast cooling already for low mass objects and the
NS with $M\simeq 1.4~M_{\odot}$ would cool faster than it is available
by modern data.
\begin{figure}[htb]
\psfig{figure=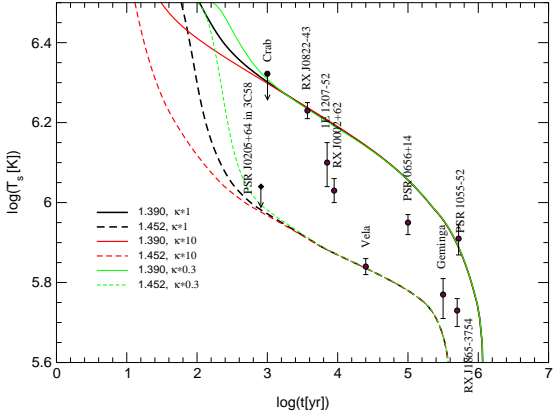,width=0.5\textwidth,angle=-90} \caption{
The influence  of a change of the heat conductivity on the scenario of
Fig. \ref{fig16}.
\label{fig18}
}
\end{figure}

Fig. \ref{fig18} demonstrates the sensitivity of the curves presented in
Fig. \ref{fig16} to the variation of $\kappa_n$.
We scale $\kappa_n$ by the factor $\zeta =10$ and $\zeta =0.3$.
We see that both increasing and decreasing of the thermal
conductivity $\kappa_n$ does not change the picture as the whole as well as the
conclusion drawn above.
As in the case when medium effects are suppressed, see Fig. \ref{fig7},
the effect of the transport is relevant for first $10^3~$yr.

\begin{figure}[htb]
\psfig{figure=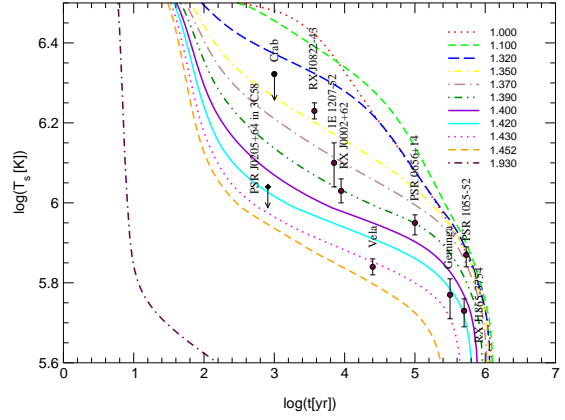,width=0.5\textwidth,angle=-90}
\caption{Same as Fig. \ref{fig16} but with suppressed proton gap
by a factor 0.5.
\label{fig20}
}
\end{figure}

\begin{figure}[htb]
\psfig{figure=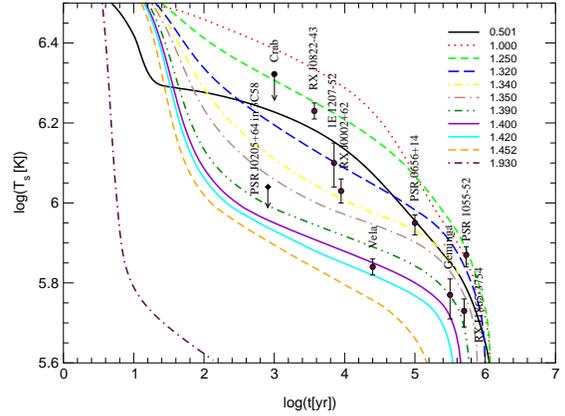,width=0.5\textwidth,angle=-90}
\caption{Same as Fig. \ref{fig16} but with $1S_0$ neutron gap suppressed by
a factor $0.5$ and $1S_0$ proton gap by a factor $0.2$.
The neutron $3P_2$ gap is remained to be
suppressed by $0.1$.
\label{fig21}
}
\end{figure}

\begin{figure}[htb]
\psfig{figure=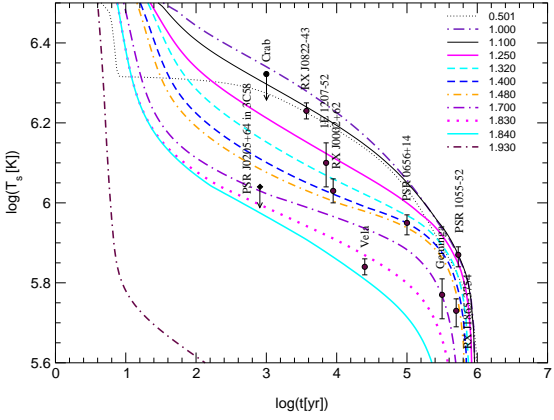,width=0.5\textwidth,angle=-90}
\caption{
Same as in Fig. \ref{fig12}  using the gaps of \cite{TT04}, see
Fig. \ref{fig5},
with additional suppression of the $3P_2$ gap by a factor 10.
\label{fig15}
}
\end{figure}

\begin{figure}[htb]
\psfig{figure=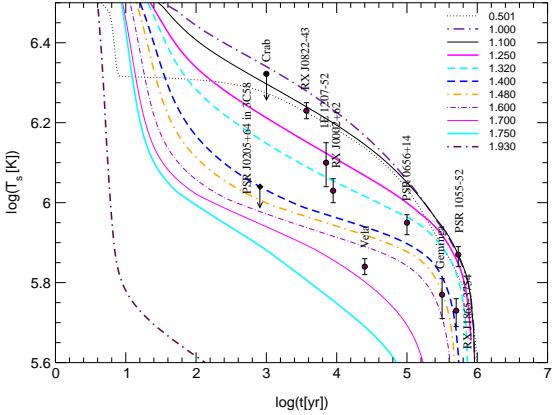,width=0.5\textwidth,angle=-90}
\caption{
Same as in Fig. \ref{fig15} but including pion condensation.
\label{fig15a}
}
\end{figure}

\begin{figure}[htb]
\psfig{figure=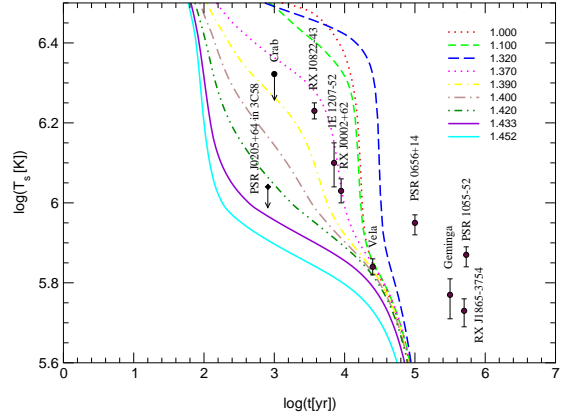,width=0.5\textwidth,angle=-90}
\caption{Same as Fig. \ref{fig16}
without suppression of $3P_2$ neutron pairing gap.
\label{fig19}}
\end{figure}

\begin{figure}[htb]
\psfig{figure=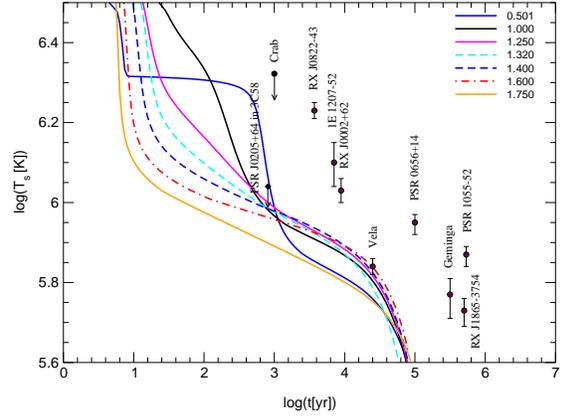,width=0.5\textwidth,angle=-90}
\caption{Same as Fig. \ref{fig15a}
without suppression of $3P_2$ neutron pairing gap.
\label{fig19a}}
\end{figure}

Figs. \ref{fig20} - \ref{fig19a} show the dependence of the results on
different variations of the gaps,
their absolute values and the density dependencies.

We remind the reader that the proton gap is artificially enhanced
by \cite{YGKLP03}. 
Fig. \ref{fig20} presents the same as Fig. \ref{fig16}, but now we suppress
the proton gap by a factor $0.5$. An appropriate fit is achieved.  
In Fig. \ref{fig21} we further suppress the proton gap (now by factor
$0.2$) and we also suppress the $1S_0$ neutron gap by factor $0.5$
simulating the medium effects in gaps. 
We again obtain an appropriate overall fit of the data. 
The NS masses covering available data are $1.33\div 1.44~M_{\odot}$ in
case of Fig. \ref{fig20} and $1.23\div 1.42~M_{\odot}$ in case of 
Fig. \ref{fig21}.
However we have checked that scaling of all gaps shown  by thick lines
in Fig. \ref{fig5} by the factor $0.1$ already does not allow for the 
appropriate fit of the data.
Thus we demonstrated important role played by all three types of the 
pairing: $1S_0$ $nn$ and $pp$ and $3P_2$ $nn$.

In Fig. \ref{fig15} we include the same medium effects as those in 
Fig. \ref{fig12} but we use the gaps according to \cite{TT04}, see 
thin lines in Fig. \ref{fig5}, additionally suppressing $3P_2$ gap 
by a factor 10. 
We see that with HHJ EoS and the given choice of the gaps we obtain 
{\em the best fit of the  whole set of the data.}
Thus the pion softening and appropriate density dependence of the 
gaps are quite sufficient to explain the modern NS cooling  data.
The data are covered by NS in wide mass interval $0.5\div 1.84~M_{\odot}$.
For masses $1.0\div 1.84~M_{\odot}$ the picture is quite regular.  
Less massive stars cool slower, more massive stars cool faster.

Fig. \ref{fig15a} presents the same as Fig. \ref{fig15} but now including
the pion condensation. 
The picture as the whole remains the same.
Only cooling of massive stars is touched that slightly narrows the 
appropriate NS mass interval from $1.0\div 1.75~M_{\odot}$ to  
$1.0\div 1.70~M_{\odot}$.

Fig. \ref{fig19} and  Fig. \ref{fig19a} present the same as  
Fig. \ref{fig16} and Fig. \ref{fig15a}, but now we take off
the suppression factor $0.1$ for the $3P_2$ neutron gap.  
From   Fig. \ref{fig19} we see that 2 {\em 'slow cooling"} 
and 2 {\em "intermediate cooling"} points related to old NS are not explained.
In case of   Fig. \ref{fig19a} only two lowest data points are explained.
We again support the statement that  {\em the suppression of the
$3P_2$ gap is indeed called for by the cooling data. }

\section{Conclusion}

We have shown that the most up-to-date observed NS cooling data
can well be explained.
Actually we deal with a many-parametric problem that allows for 
a variation of many quantities. 
Besides the data points might be partially shifted from their
positions  shown in figures due to a number of uncertainties
and assumptions used at their analysis.
By above figures we have illustrated different possibilities discriminating
more probable explanations from less probable ones.
We elaborated the example of the probably most realistic and 
microscopically supported EoS of the $V18+\delta v +UIX^*$.
Actually we used a simple parameterization of this EoS suggested 
by \cite{HJ99} (HHJ).
In this EoS the DU process does not show up to $M\simeq 1.839
M_{\odot}$. 
Thus within this model the {\em ``Standard + DU''}
scenario would demonstrate that the majority of experimentally  
measured cooling points relate to very massive NS, with 
$M\geq 1.84 ~M_{\odot}$.
From our point of view such a scenario seems unrealistic.

We exploit in-medium effects in the calculation of the emissivities 
and the pairing gaps, and, that is less important, in specific heat 
and heat conductivity.
In general, medium effects result in a significant suppression of the
superfluid gaps, especially of the $3P_2$ $nn$ gap,
and they enhance the cooling  rates of the MMU and the MNB
processes through 
{\em the pion softening effect with the increase of the density.}
Without the latter effect the {\em ``Standard + PU''} scenario 
suffers from an internal inconsistency. 
Pion condensation cannot take place without preliminary
softening of the pion mode at lower densities.
And v.s., recent argumentation for the pion condensation, cf.
\cite{APR98,SST99}, further motivates the presence of
a precursor pion softening, details see in \cite{MSTV90}. 
Pion condensation at $n\gsim 3n_0$ does not contradict the data
but the data can be explained also without pion condensation and 
other so called "exotics" (KU, HDU, etc) but with the pion softening.
Based on the works \cite{SVSWW97,BGV01} we have further improved our code.
We showed that the results are sensitive to the values and the density
dependencies of the nucleon superfluid gaps. 
{\em The strong suppression of the $3P_2$ gap} motivated by
theoretical evaluations that incorporate medium effects is indeed  
required for the fit of the data.
We show that all three groups of points {\em ``slow cooling''},
{\em ``intermediate cooling''} and {\em ``rapid cooling''}
are now well explained on the basis of the
{\em ``Nuclear medium cooling  scenario'' demonstrated here, where
an important r\^ole is played by in-medium effects}, cf. \cite{V00}.

We may draw the following main conclusions.

i) The normal matter assumption (see Figs. \ref{fig6} - \ref{fig10})
seems rather unrealistic, as by itself, as in relation to the cooling data.
One could explain the data but at the price of ignoring of medium effects in
MMU and MNB. Then the {\em ``intermediate cooling''} points can
only be explained by very low NS masses (as $0.5~M_{\odot}$) or, together with
{\em ``rapid cooling''} points, by the very high NS masses $M>1.839~M_{\odot}$,
which allow for the DU process. 
In the latter case the mass window separating the 
{\em ``intermediate cooling''} and {\em ``rapid cooling''} is very narrow.
Above price seems us too high and we  drop such a scenario. 
The superfluidity is called for by the data.

ii) Including superfluid gaps we see,
in agreement with recent microscopic findings,  that $3P_2$ neutron gap
should be as small as $\lsim 10$~keV.
Otherwise one cannot explain at least several old objects 
(see Figs. \ref{fig19}, \ref{fig19a}).
Proton and $1S_0$ neutron gaps also might be suppressed by factors
of order of several, as it is demonstrated by Figs. \ref{fig20}, 
\ref{fig21}, \ref{fig15}, \ref{fig15a}, but the suppression
by factor $\gsim 10$ is not already permitted.

iii) Medium effects associated with the pion softening are called for by the
data. As the result of the pion softening the pion condensation may occur for
$n\geq n_c^{\rm PU}$ ($n\geq 3n_0$ in our model).
Its appearance does not contradict to the data (see Fig. \ref{fig16}) but also
the data are well described, if the softening effect is rather
saturated  (see Fig. \ref{fig12}) with increase of the density 
(as demonstrated by curves 1a, 1b in Fig. \ref{fig1}).
At the same time the critical densities for the efficient DU-like
exotic processes  (such as PU on charged pion)
should not be as small as $<2.5~n_0$, if the given HHJ EoS is indeed
correct. 
Otherwise one would get too rapid cooling of the NS of the typical 
$1.4~M_{\odot}$ mass.
This also means that the proper DU threshold density can't be too low
that puts restrictions on the density dependence of the symmetry energy.  
Both  statements might be important in the discussion of the heavy ion
collision experiments.

iv) We demonstrated a regular mass dependence: for the NS masses 
$M\gsim 1.0~M_{\odot}$ less massive NS cool slower, more massive 
NS cool faster.

As we have mentioned, for the sake of simplicity
we did not include the possibility of the  hyperonization
and other possibilities, like kaon condensation and fermion condensation,
which may stimulate a more rapid cooling, being
working in a line with the pion condensation. We did not include possible
quark effects. The latter need a special treatment.
The possibility of the color
superconductivity in dense NS interiors opens a number interesting
possibilities like the so called  two-flavor color superconductivity
(2SC) phase, color-flavor-locking (CFL) phase, color-spin-locking (CSL) phase.
Their cooling is essentially
different. We will return to this discussion in the nearest future.

\begin{acknowledgements}
H.G. and D.V. acknowledge the hospitality and support of Rostock University.
The work of  H.G.  has been supported in part by the
Virtual Institute of the Helmholtz Association under grant No. VH-VI-041,
that of D.V. has been supported in part by DFG (project 436 RUS 17/117/03).
We thank E.E. Kolomeitsev for the help and discussions and S. Popov for the
discussions of the results.
\end{acknowledgements}

\end{document}